\documentclass[useAMS,usenatbib,a4paper,referee]{mn2e}
\usepackage{graphicx}
\usepackage{amsmath}
\usepackage{txfonts}

\def\aap{A\&A}
\def\apjl{ApJ}

\def\apjs{ApJS}

\title[Decaying Dark Matter]{Decaying Dark Matter \\ and the Deficit of Dwarf Haloes}
\author[Majd Abdelqader and Fulvio Melia]{Majd Abdelqader$^{1}$\thanks{E-mail:
majd@physics.arizona.edu} and Fulvio Melia$^{2}$\thanks{Sir Thomas Lyle Fellow 
and Miegunyah Fellow, E-mail: melia@as.arizona.edu}
\\
\null$^{1}$ Department of Physics, The University of Arizona, Tucson, Arizona 85721, USA
\\
\null$^{2}$ Department of Physics and Steward Observatory, The University 
of Arizona, Tucson, Arizona 85721, USA}

\date{Submitted to MNRAS 2008 January 23}

\begin{document}
\maketitle
\label{firstpage}
\setcounter{figure}{0}

\begin{abstract}
The hierarchical clustering inherent in $\Lambda$-CDM cosmology
seems to produce many of the observed characteristics of 
large-scale structure. But some glaring problems still remain,
including the over-prediction (by a factor 10) of the number of 
dwarf galaxies within the virialized population of the local group.
Several secondary effects have already been proposed to resolve
this problem. It is still not clear, however, whether the principal
solution rests with astrophysical processes, such as early feedback
from supernovae, or possibly with as yet undetermined properties of
the dark matter itself. In this paper, we carry out a detailed
calculation of the dwarf halo evolution incorporating the effects
of a hypothesized dark-matter decay, $D\rightarrow D^\prime+l$,
where $D$ is the unstable particle, $D^\prime$ is the more massive
daughter particle and $l$ is the other, lighter (or possibly massless) 
daughter particle. This process preferentially heats the smaller 
haloes, expanding them during their evolution and reducing their 
present-day circular velocity. We find that this mechanism can 
account very well for the factor 4 deficit in the observed number 
of systems with velocity 10--20 km s$^{-1}$ compared to those 
predicted by the numerical simulations, if $\Delta m/ m_{D^\prime}
\sim 5-7\times 10^{-5}$, where $\Delta m$ is the mass difference 
between the initial and final states. The corresponding lifetime
$\tau$ cannot be longer than $\sim 30$ Gyr, but may be as short
as just a few Gyr.
\end{abstract}


\section{Introduction}

Observations of the cosmic microwave background (CMB) radiation with the 
Wilkinson Microwave Anisotropy Probe (WMAP) have facilitated the precision 
measurement of several cosmological parameters (Bennett et al. 2003, Spergel 
et al. 2003), including the mass-energy density of the Universe, $\Omega$, 
which appears to be close (if not equal) to its critical value. Baryons 
contribute only about 4\% of this; the rest is presumably in the form of 
dark matter (DM; roughly 22\%) and dark energy ($\sim$74\%). In addition, 
WMAP's detection of early reionization also rules out the presence of a 
warm DM, so the non-baryonic component must be cold (CDM). Together with 
earlier observations by other finer scale CMB experiments, and with the 
Hubble Key Project (Mould et al. 2000), which provided the unprecedentedly 
accurate value $H=71\pm6$ km s$^{-1}$ of the Hubble constant, this 
combined body of work has lead to a consensus that the Universe is 
adequately described by the so-called flat $\Lambda$-CDM standard 
model, in which dark energy is the manifestation of a cosmological 
constant $\Lambda$. 

The existence of dark energy has been confirmed through the analysis 
of Type Ia supernova data (Riess et al. 1998; Perlmutter et al. 1999),
providing even stronger evidence that this component of $\Omega$ has 
negative pressure leading to an acceleration of the Universe in the
current epoc. To be fair, however, it is not yet entirely clear whether 
these results require a cosmology with a true cosmological constant, 
in which matter and radiation become dominant looking back towards 
redshifts $z>$2--3, or whether the current acceleration might be 
due to a so-called scaling solution, in which the dark energy density 
scales with matter, and affects the formation of structure even at 
early times (see Melia 2007, 2008, and references cited therein). 

Very little is known about dark matter, and almost nothing is 
understood about dark energy. Their nature is one of the biggest
mysteries in contemporary physics. Yet their influence is evidently
quite important in the formation of large scale structure (LSS).
The hierarchical clustering inherent in $\Lambda$-CDM seems to
produce many of the characteristics observed in the local (e.g.,
Murali et al. 2002; Abadi et al. 2003) and high-redshift (e.g., 
Springel, Frenk, and White 2006) Universe. In hierarchical models,
smaller dark matter haloes on average collapse earlier than larger
ones, when the density of the universe was higher (e.g., Kravtsov
et al. 1998). The current mass function, however, is determined
not only by the halo formation history, but also by their merger
rate which, over time, tends to deplete the dwarf-galaxy end
of the distribution.

But some glaring problems still remain. Numerical simulations 
involving collisionless CDM predict dark haloes with steep cusps 
in their centre (Navarro et al. 1996), whereas most of the observed 
rotation curves of dwarf galaxies and low surface brightness galaxies 
indicate constant density cores (see de Blok et al. 2003, and 
references cited therein). A related problem is the over-prediction 
of the number of dwarf galaxies within the virialized population 
of the local group. CDM simulations over-predict the number of 
satellite galaxies orbiting a Milky Way-sized galaxy by a factor 
of 10 (Klypin et al. 1999, Moore et al. 1999, Diemand et al. 2007, and Simon \& Geha 2007). Generally 
speaking, both of these problems may be described as a $\Lambda$-CDM 
prediction of too much power on small scales.

Since in hierarchical models smaller galaxies merge to make larger 
ones, the number of remaining dwarf haloes is an important diagnostic 
to test both the hierarchical picture and the process of halo 
condensation in the evolving universe. Thus, the dwarf-halo
deficit may be taken as an indicator of new physics associated
with dark matter (and/or dark energy). For example, the small-scale 
power can be reduced by substituting warm dark matter for CDM. But
as we have seen, WMAP observations have already ruled this 
possibility out.

Other simple attempts to fix the dwarf-halo deficit problem are not 
well motivated physically. Some involve altering the fundamental
nature of dark matter by introducing self-interaction, or
annihilation (see, e.g., Spergel and Steinhardt 2000; Kaplinghat, 
Knox, and Turner 2000; and Giraud 2001, among others). Without a 
proper indication from physical considerations, all of these models 
contain free parameters tunable to fit the observations. For example, 
the self-interacting dark matter scenario rests on the viability of 
a huge velocity-dependent cross section. Unfortunately, the implied
large interaction rate is inconsistent with most weakly interacting, 
massive particle and axion theories (see, e.g., Hennawi and Ostriker 
2002).

Of course, the dwarf-halo deficit may be due to reasons other than
DM physics. Many of them may be invisible because they contain a very 
small amount of luminous matter, either because of early feedback from 
supernovae (Dekel and Silk 1986; Mac Low and Ferrara 1999), or because 
their baryonic gas was heated by the intergalactic ionizing background 
radiation (Rees 1986; Barkana and Loeb 1999). Others may have turned 
into high-velocity clouds in the Local Group (see, e.g., Blitz et al. 
1999). The deficit may not even be real if the Universe is actually 
older than its current inferred age, which in reality is only the 
light-travel time to the cosmic horizon rather than the Big Bang 
(Melia 2007, 2008). In such a scenario, the dwarf haloes would have
had more time to merge, depleting the lower end of the mass function 
and bringing it into better alignment with the observations.

In any case, it is still too early to tell whether or not the discrepancy 
between the cosmological simulations and observations really indicates a major 
problem for hierarchical models in $\Lambda$-CDM. Several of the effects
we have listed here may resolve at least part of the deficit problem. However,
given that this is still an open question, observations of the halo mass
function also allow us to explore non-astrophysical reasons for the 
discrepancy, with the goal of learning more about the nature of dark matter.

Our focus in this paper is the suggestion that DM particles may be
unstable to decay (see, e.g., Davis et al. 1981; Turner et al. 
1984; Turner 1985; Dicus and Teplitz 1986; Dekel and Piran 1987; Sciama 
1990; Cen 2001a; S\'anchez-Salcedo 2003). The impact of the interactions 
we describe above and/or decays is almost always to provide a mass-dependent 
expansion of the cusps and haloes to lower the core density and to reduce 
the number of small galaxies. However, attempts to couple these ideas to 
particle physics have been few and ambiguous, partly because these have 
been concerned more with global effects, rather than aimed at finding 
specific particle properties that may be consistent with the requirements 
to fix the problem. Our goal in this paper is to begin a more careful 
search for the properties DM particles must have in order to account 
for the deficit of dwarf haloes, if in the end their decay is indeed 
responsible for the observed effect.

Our approach here is closest in spirit to the work of Cen (2001a) and
S\'anchez-Salcedo (2003), though their papers had different goals.
Cen's (2001a) primary interest was to demonstrate that if DM particles 
decay and become relativistic, they escape the virialized halo, whose
remaining energy then exceeds that required to sustain virial equilibrium
and forces it to expand. His suggestion was that the overproduction of 
dwarf haloes may be solved not by removing them, but by modifying them 
into failed, dark galaxies, in which star formation has been quenched 
due to the effects of evaporation and expansion. This is an intriguing 
idea, though not yet sufficiently developed to provide a useful probe 
into the properties of the particles themselves.

S\'anchez-Salcedo's (2003) goal was to mitigate the disparity between the
very steep central cusps in dark haloes of dwarf galaxies predicted
by $\Lambda$-CDM and the relatively flat distributions actually seen
in these systems. He demonstrated that if DM decays into a relativistic, 
nonradiative light particle, plus a stable massive particle with some 
recoil velocity in the center-of-mass frame, the former escapes the 
bound system while the latter remains with an energy exceeding that of 
the parent, forcing the halo to expand. 

In this paper, we introduce several new aspects of the DM-decay scenario,
including the time-dependent and mass-dependent halo formation and
hierarchical-merger rates in order to more accurately gauge the impact of
heating on the present-day circular velocity distribution. There are
too many aspects of the DM decay to consider in just one set of
calculations, so we here restrict our attention to cases in which at
least one of the decay products remains within the halo, maintaining
its mass, though heating it with the liberated energy. Other regions
of the DM particle phase space will be explored elsewhere. 

In the next section, we summarize the circular-velocity data and demonstrate 
the nature of the dwarf-halo deficit problem. Then in \S 3 we describe a 
technique for following the mass-dependent formation and destruction of 
haloes as the universe evolves. In \S 4 we describe the DM-particle decays 
and how we incorporate the impact of this process into our calculational 
algorithm. We present the results of our calculations in \S 5, and discuss 
them in \S 6.

\section{The Observed Circular-Velocity Distribution of Dwarf haloes}

The dwarf-galaxy deficit was first quantified when the observed number 
of dwarf galaxies in the local group was compared to high-resolution 
cosmological simulations of Klypin et al. (1999) and Moore et al. (1999), 
under the reasonable assumption that the local group is an adequate 
representation of what is happening throughout the cosmos. If we assume 
that each small dark matter halo contains a dwarf galaxy, then there is 
a substantial discrepancy between theory and observation. At the time of 
these simulations, there were only 13 known satellites of The Galaxy, 
while both simulations predicted roughly 10 times that number of 
satellites for a Milky Way-sized halo (Klypin et al. 1999, and Moore 
et al. 1999). 

In the last several years, the number of pertinent observations has
increased substantially, and new cosmological simulations have been
completed with substantially higher resolution than the original
calculations. The Sloan Digital Sky Survey (SDSS hereafter) has 
uncovered 10 more Milky Way companions (Belokurov et al. 2007). 
The discrepancy between the latest observations and the most
recent simulations still exists, however, and it appears to be 
a function of mass. In the range of primary relevance to this paper 
(i.e., 10--20 km s$^{-1}$), the disparity is a factor 4 when compared 
to the latest high resolution N-body simulation known as Via Lactea 
(Diemand et al. 2007), even after weighting the new dwarf galaxies by 
a factor of 5 to account for the limited coverage of the SDSS; at 6 
km s$^{-1}$, the discrepancy increases to a factor 10 (Simon \& Geha 2007).

\section{Formation of Bound Objects in the Hierarchical Clustering Scenario}

For reasons that will soon become apparent, the impact of decaying dark 
matter on the evolution of haloes depends on their formation history. For 
simplicity, we here use a semi-analytical approach that describes the 
halo formation rate as a function of mass and time, optimized to 
reproduce numerical simulations of structure formation. A good starting
point is the Press-Schechter mass function (hereafter PS), or one of its 
modified forms, which is a reasonable representation of the overall halo 
distribution resulting from these numerical simulations (Press and
Schechter 1974). However, PS is a number density that combines both the 
formation and merger histories of the haloes, so it
does not provide their formation rate explicitly. None the less, a formation 
rate can be extracted from the PS formalism by taking the comoving time 
derivative of the mass function, identifying the term corresponding to the 
formation rate, and multiplying it by the survivability probability. 
This procedure subtracts the halo destruction rate, and is
necessary for our purpose since we only want to consider haloes that 
formed in the past and survived to the present without merging with
others to form even bigger structures (see, e.g., Sasaki 1994; 
Kitayama \& Suto 1996). The formation rate can be written as (Sasaki 1994)
\begin{equation}
\dot{N}_{form}(M,t) \;dM \;dt = \frac{1}{a(t)} \frac{d a(t)}{dt}N_{PS}
(M,t)\frac{\delta^{2}_{c}(t)}{\sigma^{2}(M)}\;dM \;dt \ ,
\end{equation}

\noindent where $M$ is the mass of the formed gravitational structure, $t$ is the 
formation time measured in comoving coordinates starting from zero at the 
Big Bang, $a(t)$ is the cosmological scale factor normalized to unity at 
the present epoch $t_{0}$, $N_{PS}(M,t)$ is the Press-Schechter mass 
function, $\delta_{c}(t)=\frac{\delta_c}{a(t)}$ is the critical density 
threshold for a spherical perturbation to collapse by time $t$ ($\delta_c 
\simeq 1.69$ for $\Omega_0 =1$), and $\sigma(M)$ is the rms density 
fluctuation smoothed over a region of mass $M$. However, we are interested 
in haloes surviving to the present epoch, thus the above formation rate 
function must be multiplied by the probability $p(t_1,t_2)$ that an object 
which exists at $t_1$ remains at $t_2$ without merging, which is given by 
$p(t_1,t_2)={a(t_1)}/{a(t_2)}$ (Sasaki 1994). The formation rate distribution 
of surviving haloes at the present epoch becomes $F(M,t)\;dM\;dt=\dot{N}_{form}
(M,t)\times p(t,t_0)\;dM\;dt$, which can be written explicitly as
\begin{equation}
F(M,t)\;dM\;dt=A \frac{da(t)}{dt} \frac{1}{a(t)^3} M^{(n-1)/2} \exp 
\left[ -\tfrac{1}{2}  
\left(\tfrac{M}{M_{c,0}} \right)^{(n+3)/3} \frac{1}{a(t)^2} \right] \ ,
\label{formation}
\end{equation}

\noindent where A is a normalization constant, $n$ is the power-law spectral index 
chosen to be $-2.5$, following Klypin et al. (1999), and $M_{c,0}$ is the 
present mass scale of the knee taken to be $ 10^{15}\, M_{\odot}$.

In this paper, we adopt a flat universe $\Omega_0=1$, represented roughly
as $\Omega_{m,0}=0.3,$ and $\Omega_{\Lambda,0}=0.7$, which gives 
$a(t)= \left({0.3}/{0.7} \right)^{1/3}\, \sinh ^{2/3} \left(  1.21 \,[{{t}/{t0}]} \right)
$, normalized to unity at the present epoch $t_0=13.7$ Gyr. Figure~1 shows the formation 
rate as a function of time for two illustrative halo masses. The larger mass 
always has the smaller formation rate, which also peaks at later times,
though this difference is not very obvious from the figure as the two are 
relatively close.

\section{DM Particle Decays and their Impact on Dwarf Halo Evolution}

In this paper, we focus our attention on dark matter decay scenarios 
in which the parent particle, denoted by $D$, decays into one stable 
massive particle ($D'$) with a mass close to that of its parent, and 
one very light (possibly even massless) particle ($\ell$), following 
the notation of S\'anchez-Salcedo (2003). Due to the extreme mass 
ratio between the daughter particles, the light particle carries most 
of the energy released by the decay and becomes relativistic, escaping 
the halo. The massive daughter particle remaining in the halo has an 
(average) energy very close to, but slightly larger, than that of its parent, 
forcing the halo to expand adiabatically. Since the kinetic energy
of the light particle is much larger than its rest mass, we neglect 
the latter. To this level of approximation, the total mass of the 
halo remains unaffected by the decay because $ m_{_{D'}}\approx m_{_D}$. 

\begin{figure}
\center{\includegraphics[scale=0.65,angle=-90]{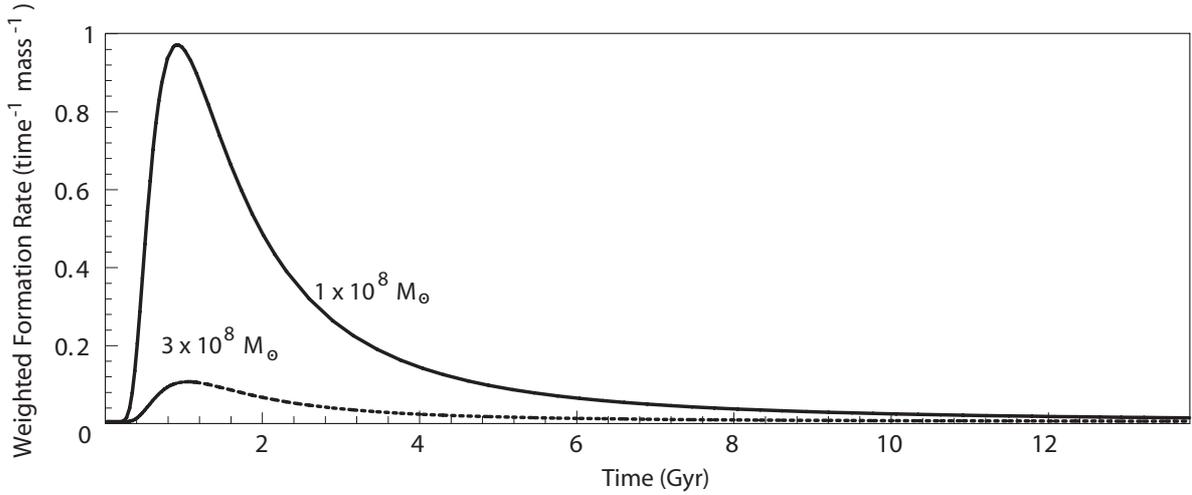}
\vspace{10pt}
\caption{The relative halo formation distribution as a function of cosmic
time for haloes surviving to the present epoch, for two illustrative
masses: $M=1\times10^8 M_{\odot}$ and $M=3\times10^8 M_{\odot}$.}}
\end{figure}

The decay is represented as $ D \rightarrow D'+\ell $. In the 
center-of-mass frame (denoted by a * superscript), i.e., in the rest 
frame of the parent particle $D$, the initial four-momentum vector 
before the decay is 
\begin{equation}
p_i^*=\big(m_{_D} c \, , \vec{0}\, \big)\;,
\end{equation}
and after the decay the final four-momentum is 
\begin{equation}
p_f^*=p_{_{D'}}^*+p_{_{\ell}}^*=\big( \ \gamma_{_{D'}}^* m_{_{D'}} c\ ,
\  -\gamma_{_{D'}}^* m_{_{D'}} v_{_{D'}}^*\ ,\ 0\ ,\ 0\ \big) + 
\big( \  E_{\ell}^*/c \ ,\ E_{\ell}^*/c \ ,\ 0\ ,\ 0\  \big)\;,
\end{equation}  
where we have put the direction of $ \vec{p_{_{\ell}}}$ along $+\hat{x}$, 
and $ \vec{p_{_{D'}}} $ along $-\hat{x}$. 

Conservation of momentum leads to the equation
\begin{equation}
\ E^{*}_{\ell}=(m^{2}_{_{D}}-m^{2}_{_{D'}})c^{2}/2m_{_D}\;.
\end{equation}
To find the particle's energy $E_{\ell} $ in the halo's rest frame, 
we need to Lorentz boost the physical quantities using the
velocity $ \vec{v}=\vec{v}_{_D} $ and the angle $ \theta^* $ between 
$\vec{v}$ and $ \vec{p_{_{\ell}}}$. This gives 
\begin{equation}
E_{\ell} =\gamma_{_D} E^*_{\ell} (1+\beta_{_D} \cos \theta^*)\;.
\end{equation}
Averaging over all solid angles, we find that
\begin{equation}
\left\langle E_{\ell} \right\rangle  = \frac{1}{4\pi} \int E_{\ell} 
\, d\Omega = \gamma_{_D} E^*_{\ell}\;.
\end{equation}

We next define the unitless parameter 
\begin{equation}
\chi \equiv \frac{\Delta m}{m_{_{D'}}}\;,
\end{equation}
the ratio between the change in rest mass $\Delta m =m_{_{D}}-m_{_{D'}}$
and the mass of the heavier daughter particle $D'$. The energy of the 
lighter (relativistic) particle, averaged over all angles $ \theta^* $, 
becomes
\begin{equation}
\left\langle E_{\ell} \right\rangle =  \gamma_{_D} m_{_{D}} 
\frac{\chi(2+\chi)}{2(1+\chi)^2}\, c^2\;.
\end{equation} 

If both particles remain in the halo, then the change in the halo's 
energy for each decay is simply $ \Delta m \, c^2$. However, the lighter 
particle escapes, so its energy must be subtracted. We find that for
this type of decay, the rate of change in the halo's energy is therefore
\begin{equation} 
\frac{dE_{halo}}{dN} =\Delta m \, c^2-\left\langle E_{\ell} 
\right\rangle =  m_{_{D}} c^2 \left( \frac{\chi}{1+\chi}- \gamma_{_D} 
\frac{\chi(2+\chi)}{2(1+\chi)^2} \right)\;,
\end{equation}
where $N$ is the number of unstable particles. But the decay rate is
\begin{equation}
\frac{dN}{dt}= \frac{d}{dt} N_0 e^{-(t+t_{f})/\tau} 
= -\frac{N_0}{\tau}e^{-(t+t_{f})/\tau}\;,
\end{equation}
where $t_{f} $ is the time at which the halo forms, $t$ is the time 
elapsed since the formation of the halo, $\tau$ is the mean lifetime 
of the parent particle $D$, and $ N_0$ is the initial number of parent 
particles at the time the universe began its expansion. Combining these
quantities, we can now find the rate at which the halo's energy changes
with time:

\begin{eqnarray}
\quad\qquad\qquad\frac{dE}{dt}=\frac{dE}{dN} \, \left|\frac{dN}{dt}\right|&=&
\frac{(N_0 m_{_{D}}) c^2 }{\tau}e^{-(t+t_{f})/\tau} \,  
\left( \frac{\chi}{1+\chi}- \gamma_{_D} \frac{\chi(2+\chi)}{2(1+\chi)^2} \right) \nonumber \\
\null&=&\frac{(M) c^2 }{\tau}e^{-(t+t_{f})/\tau} \,  
\left( \frac{\chi}{1+\chi}- \gamma_{_D} \frac{\chi(2+\chi)}{2(1+\chi)^2} \right) \;.
\label{dEdt1}
\end{eqnarray}
\vskip 0.1in
By knowing the rate of energy change, we can in principle find the change in size
of the halo by expanding it adiabatically. However, in order to do that, we need 
to know its initial density profile. At the time of formation, we will assume the 
halo has a Navarro, Frenk, \& White density profile (1997, NFW hereafter) 
\begin{equation}
\frac{\rho_{_{\mathrm{NFW}}}(r)}{\rho_{_{\mathrm{crit}}}}=
\frac{\delta_{\eta}}{(r/r_s)(1+r/r_s)^2 }\;,
\end{equation}
where $ \rho_{_{\mathrm{crit}}} $ is the critical density, $r_s$ is the scale radius 
of the NFW profile, and 
\begin{equation}
\delta_{\eta}= \frac{200}{3} \frac{\eta^3}{\ln(1+\eta)-
\eta/(1+\eta)}
\label{delta}
\end{equation}  
is a characteristic (dimensionless) density in terms of 
$\eta=r_v/r_s$ (the concentration parameter) and the virial radius $ r_v$ 
(defined as the radius of a sphere of mean interior density $200
\rho_{_{\mathrm{crit}}}$). Although the NFW profile is an analytic function, 
it cannot be readily incorporated into our semi-analytical model. First, its 
distribution function $f(r,v)$ cannot be obtained analytically, and must be 
found numerically. Furthermore, to find how the scale radius $r_s $ evolves 
with time, we need to find the halo energy as a function of $ r_s$, but 
this is not easy to do with the NFW profile. Assuming the halo is virialized, 
$ E_{tot}=\left\langle K \right\rangle+\left\langle U 
\right\rangle=-\frac{1}{2}\left\langle U \right\rangle+ \left\langle U 
\right\rangle=\frac{1}{2} \left\langle U \right\rangle$. The total potential 
energy is given as (Binney \& Tremaine 1987)
\begin{equation}
U=\frac{1}{2} \int{ \Phi ({\mathbf x}) \rho ({\mathbf x}) \, d^3  {\mathbf x}}\;,
\label{energy}
\end{equation}
where $\Phi$ is the gravitational potential.

To evaluate this integral, we need to find $\Phi ({\mathbf x})$ from 
$\rho ({\mathbf x})$, which we can do by first finding the mass 
\begin{equation}
M(r)=\int\limits_0^r{ 4\pi r'^2 \, \rho (r') \, d r'}\;.
\label{mass}
\end{equation}
For the NFW profile, 
\begin{equation}
M_{_{\mathrm{NFW}}}(r)=M\frac{\ln(1+r/r_s)-\frac{r/r_s}{1+r/r_s}}
{\ln(1+\eta)-\frac{\eta}{1+\eta}}\;,
\label{NFWmass}
\end{equation}
where $M$ is the virial mass of the halo contained inside the virial 
radius. By definition, $\nabla \Phi({\mathbf x})=- {\mathbf F({\mathbf x})}$, 
and for the simple isotropic case, $ {\mathbf F(r)} = -GM(r)/r^2$, which leads to
\begin{equation}
\Phi(r)=\int\limits_0^r{\frac{G M(r')}{r'^2}\, dr' }\;. 
\label{potential}
\end{equation}
For an NFW halo, this gives\footnote{ To get the correct energy, the potential 
$\Phi ({\mathbf x})$ must be adjusted with an additive constant such that 
$\Phi (r_v)=-GM/r_v$ since the NFW halo density must vanish for $r>r_v$.}
\begin{equation}
\Phi_{_{\mathrm{NFW}}}(r)= \frac{-GM}{r}\,\frac{\ln(1+r/r_s)-\frac{r/r_s}
{(1+\eta)}}{\ln(1+\eta)-\frac{\eta}{1+\eta}}\;.
\end{equation}
Thus, solving for $U$ in equation~(\ref{energy}), we find that 
\begin{equation}
E_{_{\mathrm{NFW}}}= -\frac{GM^2}{4r_s} \left( \frac{1-1/(1+\eta)^2-2\ln(1+\eta)/(1+\eta)}
{\left[\ln(1+\eta)-\eta/(1+\eta) \right]^2} \right)\;.
\label{nfwenergy}
\end{equation}
In this equation, the concentration parameter $\eta$ changes with $r_s$, which
makes it difficult to find an analytic expression for $ dr_s/dt $ in terms of 
$dE/dt$. 

For these reasons, it is convenient to translate the NFW profile into an
equivalent Plummer distribution, 
\begin{equation}
\rho_{_p}(r)=\left( \frac{3M}{4 \pi r_p^3} \right) 
\left( 1+ \frac{r^2}{r_p^2}\right)^{-5/2}\;,
\end{equation}
which is mathematically easier to evolve in time. In this expression, 
$r_p$ is the Plummer scale radius. A principal virtue of the Plummer 
profile is that it solves the Lane-Emden equation for a self-gravitating, 
polytropic gas sphere. The corresponding distribution function may be written
\begin{equation}
f_p(r,v)=B\left[\frac{GM}{\sqrt{r^{2}+r_p^{2}}}-\frac{1}{2}v^{2}\right]^{7/2}\;,
\label{distribution}
\end{equation}
where $B$ is a normalization constant, and it is trivial to show from 
equations~(\ref{mass}) and (\ref{potential}) that
\begin{equation}
M_p(r)=M \frac{r^3}{\left(r_p^2+r^2\right)^{(3/2)}}\;,
\end{equation}
and
\begin{equation}
\Phi_p(r)=\frac{-GM}{\sqrt{r_p^2+r^2}}\;.
\end{equation}

Thus, the total energy of a Plummer halo (from equation~\ref{energy}) is
\begin{equation}
E_p=-\frac{3\pi GM^{2}}{64r_p}\;.
\label{plummerenergy}
\end{equation}
In making the transition from an NFW profile to its corresponding Plummer
form, we use the physically reasonable criterion that two haloes should have
the same mass and energy. Therefore, equating equations~(\ref{nfwenergy}) and 
(\ref{plummerenergy}), we get
\begin{equation}
 r_p=r_s\frac{3\pi}{16}  \left(\frac{\left[\ln(1+\eta)-\eta/(1+\eta) 
\right]^2}{1-1/(1+\eta)^2-2\ln(1+\eta)/(1+\eta)} \right)\;.
\end{equation}
Following Navarro, Frenk, and White (1997), we will further assume that 
$\eta=20$ at the time of formation (for the smallest haloes), though our 
results are insensitive to its actual value. 

Figure~2 illustrates the differences in circular velocity 
($ v_{\rm circ}=\sqrt{GM(r)/r}$) for an NFW halo (with mass 
$3\times 10^8\; M_\odot$) and the corresponding equivalent
Plummer form. The Plummer halo may not fit the observed velocities 
as well as NFW, but their maximum circular velocities are almost the 
same, and since the Plummer model permits us to obtain an analytic 
solution for the halo's evolution in time, we will
use it in all our calculations under the assumption that the collective 
behavior of self-gravitating virialized haloes is similar for slightly 
different profiles. Nevertheless, we will still need to use the NFW 
profile to obtain the initial characteristics of the halo at the time 
of its formation.

\begin{figure}
\center{\includegraphics[scale=0.65,angle=-90]{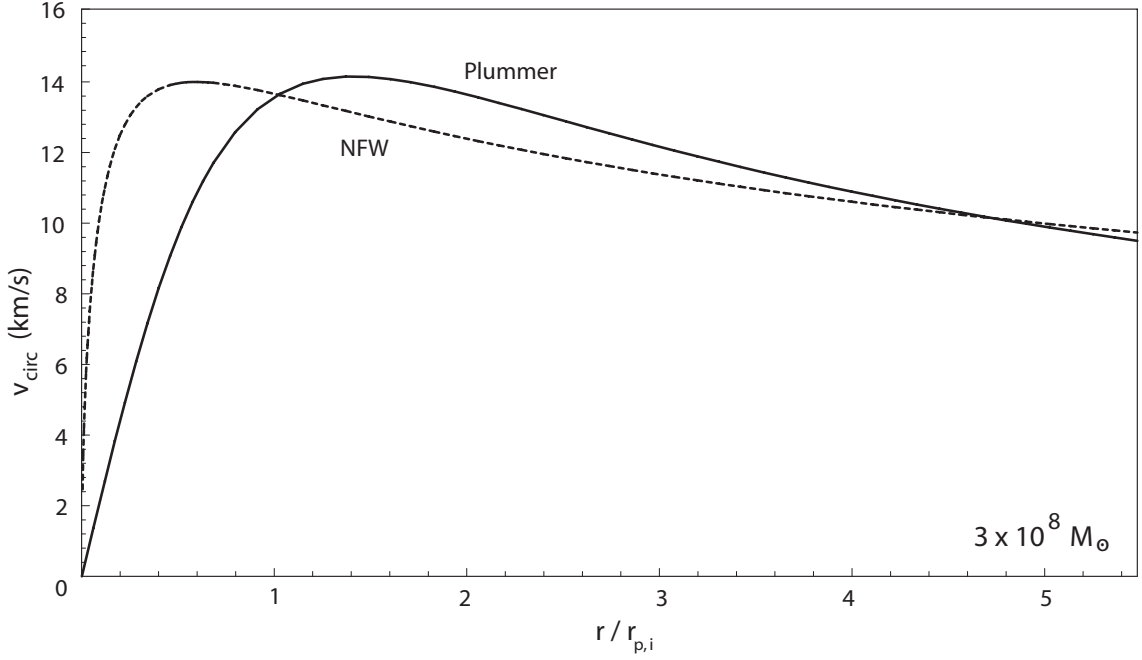}
\vspace{10pt}
\caption{Circular velocity curves of two haloes with the same mass 
($M=3\times10^8 M_{\odot}$) and the same total internal energy, but 
one with a Plummer density profile, and the other with an NFW density 
profile. Radii are in units of the initial Plummer scale radius, 
$r_{p,i}$, and circular velocities in units of kilometers per second.}}
\end{figure}

Now, using the Plummer distribution (equation~\ref{distribution}), we can 
simplify equation~(\ref{dEdt1}) by evaluating the average rate\footnote{Note
that $ \gamma_{_D}=1/\sqrt{1-v^2/c^2} \approx 1+v^2/2c^2$ since particle 
$D$ is non-relativistic.}
\begin{eqnarray}
\qquad\qquad\left\langle  \frac{dE}{dt} \right\rangle\hskip-0.1in&=&
\hskip-0.1in \int \frac{dE}{dt} 
f_p(r,v)\, d^3 {\mathbf x}\, d^3 {\mathbf v}\nonumber \\ 
\null\hskip-0.1in&=&\hskip-0.1in \frac{(M) c^2 }{\tau}
e^{-(t+t_{f})/\tau} \,  \left[ \frac{\chi}{1+\chi}- 
\left( 1+ \frac{3 \pi G M}{64 c^2 r_p} \right) 
\frac{\chi(2+\chi)}{2(1+\chi)^2} \right]\;.
\label{dEdt2}
\end{eqnarray}

\begin{figure}
\center{\includegraphics[scale=0.65,angle=-90]{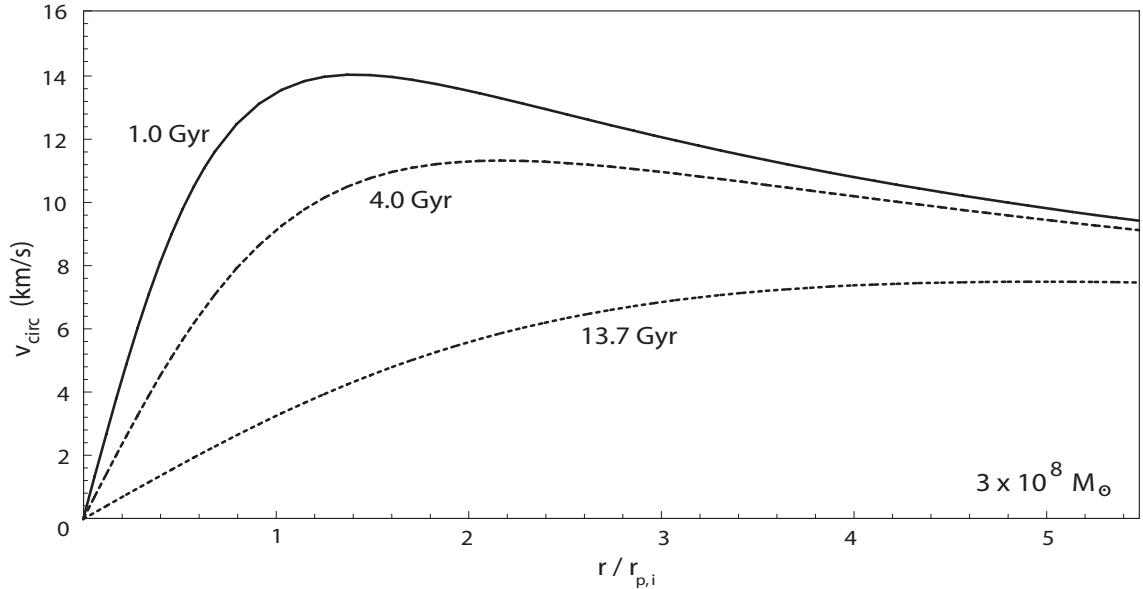}
\vspace{10pt}
\caption{Circular velocity curves at different times (indicated) of an expanding 
halo formed at $t=1 Gyr$ with mass $3\times10^8 M_{\odot}$. The expansion is due to
the decay of CDM particles with $\chi=4\times10^{-5}$ and $\tau=5$ Gyr. The radius
is in units of the initial Plummer scale radius, and circular velocities are km s$^{-1}$.}}
\end{figure}

And combining equations~(\ref{plummerenergy}) and (\ref{dEdt2}), we therefore get
\begin{eqnarray} 
\qquad\qquad\frac{dr_p}{dt}\hskip-0.1in&=&\hskip-0.1in \frac{64\, r_p^2}{3\pi G M^2} 
\frac{dE}{dt}\nonumber \\ 
\null\hskip-0.1in&=&\hskip-0.1in\frac{64\, r_p^2}{3\pi G M} \frac{c^2 }{\tau}
e^{-(t+t_{f})/\tau} \, \left[ \frac{\chi}{1+\chi}- \left( 1+ 
\frac{3 \pi G M}{64 c^2 r_p} \right) \frac{\chi(2+\chi)}{2(1+\chi)^2} \right]\;.
\end{eqnarray}

This differential equation can be solved analytically, yielding the result\vskip -0.08in
\begin{equation}
 r_p (t)=r_{p,i} \frac{3\pi G M (\chi +2) \exp  \left[-\frac{\chi (\chi+2) }
{2(\chi+1)^2}e^{-t_{f}/\tau}\right]}{64 \chi c^2r_{p,i}\exp  \left[-
\frac{\chi (\chi+2) }{2(\chi+1)^2}e^{-t_{f}/\tau}\right]+ 
\left( 3 \pi G M (\chi+2)-64 \chi c^2 r_{p,i} \right)\exp  
\left[-\frac{\chi (\chi+2) }{2(\chi+1)^2}e^{-(t+t_{f})/\tau}\right]}\;,
\label{finalresult}
\end{equation}
\vskip -0.08in\noindent
where $r_{p,i}$ is the initial Plummer scale radius at the time of formation. 
Note that the impact of DM decay on the evolution of the halo does not depend 
on the mass of the individual particles $D$ and $D'$, but rather on the 
ratio of these masses represented by the unitless parameter $\chi$, as well 
as on the mean lifetime $\tau$ of the parent particle $D$. There is an 
underlying assumption here that the halo keeps a Plummer profile throughout 
the expansion. When comparing the decay time scale, defined as $t_{decay} 
\equiv \left| {E}/{dE/dt} \right|$, to the dynamical time $t_{dyn}\equiv\sqrt{{3 \pi}/
{16 G \rho}}$ (Binney \& Tremaine 1987), we find that $t_{decay}/t_{dyn} > 10$,
implying that the halo always equilibrates to its virialized profile fast 
enough to justify the quasi-equilibrium approximation. Figure~3 illustrates 
the expansion of a halo formed at $t_{f}=1$ Gyr, showing how the maximum 
circular velocity decreases with time.

\section{Computations and Results}

To explore the global impact of our model on the halo distribution, 
we incorporate the effects of the DM decay on each individual halo 
(equation \ref{finalresult}) and its formation rate (equation 
\ref{formation}) through a series of calculations. We assess 
the consequences of this process by examining the modifications
to the distribution of maximum circular velocity ($v_{circ}$ hereafter) 
under two assumptions: (i) that at the time of formation, all haloes 
with the same mass will have the same concentration parameter. Although 
this is evidently incorrect for individual haloes, we are considering
the global behavior, for which an average concentration parameter will
suffice. Thus, in the case of no decay, all virialized haloes with the 
same mass will have the same circular velocity regardless of when they 
formed, since their concentration parameter does not change with time 
and only depends on the halo's mass; (ii) that the stellar dispersion 
velocity is proportional to the maximum circular velocity of the host 
dwarf halo.\footnote{This assumption has been extensively used to infer 
the circular velocity for elliptical galaxies and dwarf spheroidals from 
the observed stellar dispersion velocity $v_{circ}=\sqrt{3}\sigma$, 
where $\sigma$ is the observed stellar dispersion velocity (Klypin et 
al. 1999, Simon \& Geha 2007)} Thus, with DM decay, the haloes expand 
and their concentration will correspondingly change with time, which 
in turn alters the circular velocity and the observed stellar dispersion 
velocity. So haloes with different masses that formed at different times 
in the past may end up with the same circular velocity in the current epoch.

\begin{figure}
\center{\includegraphics[scale=0.8,angle=0]{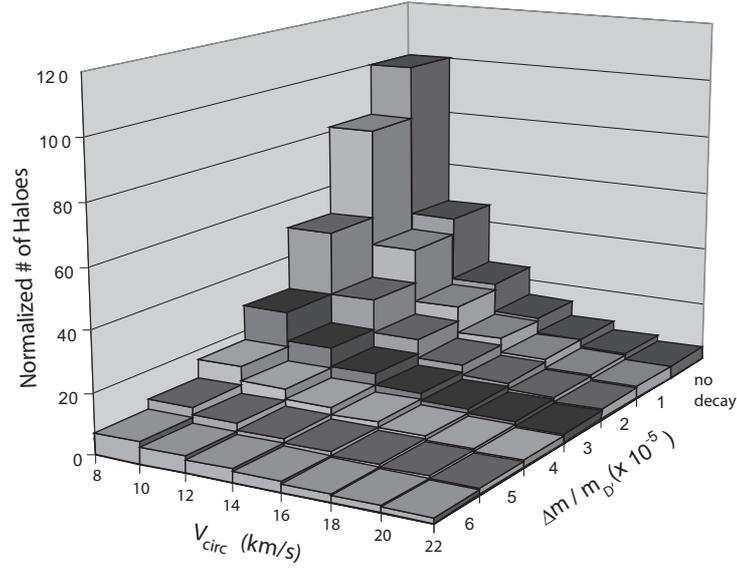}
\vspace{10pt}
\caption{Circular velocity distribution of DM haloes for a fixed mean lifetime 
$\tau= 5$ Gyr, and different values of the decay parameter $\chi\equiv
\Delta m/m_{D^\prime}$. The circular velocity distribution without decay is 
provided for comparison.}}
\end{figure}

For a given decay parameter $\chi$ and a mean lifetime $\tau$, the 
evolution of the halo still depends on when it formed and on its mass. 
For example, if the halo formed late relative to the  mean lifetime 
$\tau$, then most of the unstable DM particles will have already 
decayed by then, and the halo will therefore experience no 
significant expansion. Furthermore, according to equations 
(\ref{dEdt1}) and (\ref{plummerenergy}), $dE/dt\propto~M$, 
while $E\propto~M^2$, which means that $({dE}/{dt})/E\propto1/M$. 
This is a crucial dependence of this process on mass since it guarantees
a relatively stronger impact on the smaller haloes. For example, if a small 
halo with $M\approx10^{8}\, M_{\odot}$ experiences a significant expansion 
for a given set of decay parameters (for which, say, $({dE}/{dt})/E$ is of
order unity), then a much larger halo, e.g., with $M\approx~10^{10} M_{\odot}$, 
will be unaffected by the decay since now $({dE}/{dt})/E \approx~10^{-2}$.

Figure~4 demonstrates the impact of changing the decay parameter $\chi$ on 
the circular-velocity distribution in the range 8--22 km s$^{-1}$, given a 
fixed mean lifetime. For this calculation, the formation time distribution 
function was normalized such that without any DM decay, 100 haloes would
be produced with $v_{circ}$ between 10 and 20 km s$^{-1}$. We can see that 
the decay decreases the number of haloes with lower circular velocities 
more than those with higher ones. Furthermore, the impact of the DM decay 
on the velocity distribution increases with $\chi$, which is not surprising
given that a higher $\chi$ corresponds to a higher recoil velocity of the 
remaining particle $D'$ and, therefore, a greater expansion. It is important 
to note that our approximations cease to be valid at $\chi \approx 7\times10^{-5}$, 
because for higher values the recoil velocity of $D'$ becomes comparable to 
its escape velocity in the haloes we are considering. For such high values
of $\chi$, some of the $D'$ particles would start escaping right after the 
decay, depending on the velocity and initial position of the parent particle 
and on the angle $\theta^{*}$ that the recoil velocity makes with the velocity
of the parent particle. As a result, the mass of the halo would decrease with 
time, an effect that is not being taken into account right now.

\begin{figure}
\center{\includegraphics[scale=0.8,angle=0]{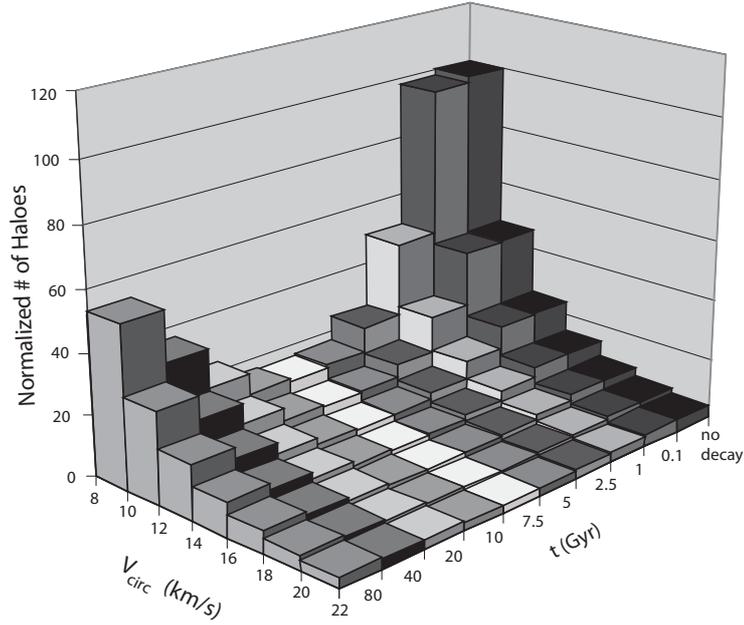}
\vspace{10pt}
\caption{Circular velocity distribution of DM haloes for a fixed decay parameter 
$\chi=4 \times 10^{-5}$, and different values of the mean lifetime $\tau$. 
The circular velocity distribution without decay is provided for comparison.}}
\end{figure}

We have also examined the impact of varying the mean lifetime $\tau$ on our
model, the results of which are summarized in figure~5, for a fixed value of 
$\chi$. For relatively small values of $\tau$ ($\approx 0.1$ Gyr), the decay 
has virtually no impact on the circular velocity distribution since most of 
the DM particles will have already decayed before the vast majority of the 
haloes formed. In these cases, the haloes experience no significant expansion. 
As $\tau$ increases, however, most of the DM particles decay after the majority
of the haloes have formed.\footnote{This implies that the dependence of our
model on $\tau$ is strongly related to the formation time distribution function. 
Thus, using a different formation function can produce a qualitatively different 
dependence on $\tau$.} The impact of DM decay reaches a maximum "effectiveness" 
in decreasing the number of haloes for $\tau\sim 10$ Gyr. As one would expect,
larger values of $\tau$ produce less significant results since most of the
unstable particles would not have decayed by the present time. 

\begin{figure}
\center{\includegraphics[scale=0.65,angle=0]{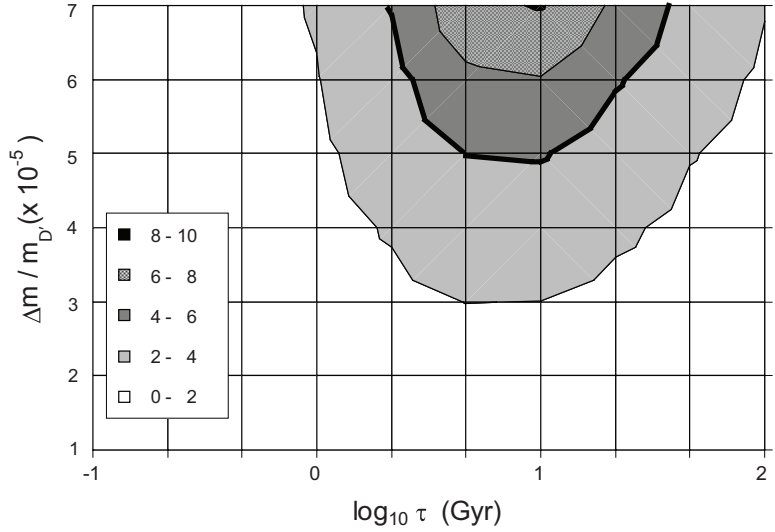}
\vspace{10pt}
\caption{Contour plot representing the effectiveness of the DM decay in 
reducing the number of dwarf haloes with $v_{circ}$ between 10 and 20 km s$^{-1}$, 
given a set of decay parameters $\chi\equiv\Delta m/m_{D^\prime}$ and $\tau$. 
The shaded regions represent the ratio of the number of haloes predicted without 
decay (for $v_{circ}$ between 10 and 20 km s$^{-1}$) to that obtained with DM decay.}}
\end{figure}

The full dependence of the circular-velocity distribution on DM decay is
shown in figure~6, which includes the effects of both $\chi$ and $\tau$. 
The contour levels represent the effectiveness of the decay in decreasing 
the number of dwarf haloes within the velocity range 10--20 km s$^{-1}$. 
The various shaded regions represent the expected number of haloes without 
decay divided by the corresponding number when expansion occurs with a
given set of parameters $\chi$ and $\tau$. 

Besides adjusting the predicted circular velocity distribution to bring it
in line with observations, the most important observational signature of 
DM decay is the dependence of the concentration parameter of dwarf galaxies 
on redshift. At relatively large redshifts, the haloes would have formed 
recently and a smaller fraction of the DM particles would have decayed,
so neither the concentration parameter, nor the stellar dispersion velocity,
would have been influenced greatly. This is to be contrasted with what would 
have happened to haloes observed in the current epoch. Therefore, distant dwarf 
haloes at large redshifts would be expected to be brighter (on average) and to 
have higher dispersion velocities compared to their nearby (lower redshift) 
counterparts with similar masses.

\section{Conclusion}
We have shown that the decay of unstable CDM particles can fully account for 
the deficit of dwarf galaxies in the local group, and have identified some of 
the particle properties required to achieve this result. In figure~6, the
lifetime $\tau$ and energy conversion fraction $\chi$ must have values
consistent with the thick black line between the two gray regions 
in order to reduce the number of dwarf haloes
within the range of velocities 10--20 km s$^{-1}$ by a factor $\sim$ 4, 
in agreement with the observations (Simon \& Geha 2007). 
Broadly speaking, the DM decay model works very well as long as $\chi
\sim 5-7\times 10^{-5}$. The lifetime cannot be longer than $\sim 30$ Gyr,
and may be as short as only a few Gyr, depending on the precise value
of $\chi$. Note, however, that we have here restricted our analysis to 
cases in which the more massive decay particle $D^\prime$ remains bound 
to the halo. We may find other regions of $\chi-\tau$ phase space that 
produce reasonable results when this restriction is removed.

We emphasize that this model can reduce power on small scales consistent 
with the observations without altering the number of Milky Way-sized galaxies. 
Very importantly, we have shown that although the expansion produced by these 
decays changes the circular velocity distribution, it does not change the halo 
mass function, at least not directly. It is beyond the scope of the present paper 
to seek the ultimate fate of dwarf haloes expanding to circular velocities below 
10 km s$^{-1}$, which remains an open question, though several possibilities have 
been proposed by Cen (2001a, 2001b). 

It should be pointed out, however, that in addition to expanding (preferentially) 
the smaller haloes, and thereby reducing their measurable velocity dispersion, 
DM decay would also have the effect of speeding up their rate of evaporation 
within the Milky Way's tidal field. In this way, dwarf haloes would be removed 
from the overall velocity distribution, not only due to their migration in velocity 
space towards the low end, but would also be removed entirely due to evaporation. 
This effect would not be evident with the larger haloes, for which DM decay
would have little impact on their condensation (and hence on their velocity profiles).

Interestingly, some limits on DM-decay models have already been established
based on limits provided by the cosmic $\gamma$-ray background. Although it is 
beyond the scope of the present paper to consider the implications of our work on
all possible DM scenarios, it is useful to see how coupling our astrophysically-motivated 
simulations to the various particle physics proposals could develop in the future. 
For example, in their consideration of WIMPs decaying to Kaluza-Klein gravitons 
and gravitinos, Cembranos et al. (2007) demonstrated that both the energy spectrum 
and flux of the observed diffuse MeV $\gamma$-ray excess may be explained by 
decaying DM with $\sim$MeV mass splittings. In this picture, a decay timescale 
of 10 Gyr would require a mass splitting of $\sim 10$ MeV, for which the unstable 
DM particle would then have a mass $\sim 1$ TeV within the context of our model. 
In a second example, Kribs and Rothstein (1997) placed bounds on long-lived
primordial relics using measurements of the diffuse $\gamma$-ray spectrum from
EGRET and COMPTEL. They concluded that relics decaying predominantly through
radiative channels are excluded for lifetimes between 
$10^{5}$ and $10^{15}$ years. Since the DM decay timescale in
our model fits within this range, the radiative decay of
relics such as these could not resolve the dwarf spheroidal problem.

We have kept our analysis semi-analytical in order to better gauge the impact
of our assumptions and chosen parameter values. Of course, to get a more 
accurate set of results, one should couple the properties of decaying DM 
particles with a more realistic N-body simulation. We intend to carry
out such a calculation in the near future and will report the results
elsewhere.

\vspace*{-0.3cm}
\section*{Acknowledgments}

This research was partially supported by NSF grant 0402502 at the
University of Arizona, and a Miegunyah Fellowship at the University
of Melbourne. We are very grateful to Romeel Dav\'e for very helpful 
discussions. Part of this work was carried out at the Center for 
Particle Astrophysics and Cosmology in Paris.

\end{document}